\documentclass[secnumarabic,superscriptaddress,amssymb, aps, onecolumn, prl]{revtex4-2}

\usepackage{amsmath}
\usepackage{mathrsfs}  
\usepackage{mathtools} 
\usepackage{bbold}    
\usepackage{braket}    
\usepackage{graphicx}				
\usepackage{amssymb}

\usepackage{gensymb}   

\usepackage{graphicx}      	
\usepackage{wrapfig}
\usepackage{multirow}       
\usepackage{here} 			
\usepackage{url}		
\usepackage[plainpages=false,pdfpagelabels=true, linkcolor=cyan,breaklinks,colorlinks=true]{hyperref}  
\usepackage{soul}                   
\usepackage[dvipsnames]{xcolor}		

\usepackage{xspace}
\usepackage{booktabs}    
\usepackage{titlesec}
\newcommand{\bcao}{BaCo$_2$(AsO$_4$)$_2$\xspace}
\newcommand{\bco}{$B_{\mathrm{c}1}$\xspace}
\newcommand{\xxz}{XXZ-$J_1$-$J_3$\xspace}
\newcommand{\bct}{$B_{\mathrm{c}2}$\xspace}
\begin{document}

\title{Quantum Fluctuations Suppress the Critical Fields in BaCo$_2$(AsO$_4$)$_2$}

\author{Shiva Safari}
\affiliation{Institute of Science and Technology Austria, 3400 Klosterneuburg, Austria}

\author{William Bateman-Hemphill}
\affiliation{Department of Physics, University of Toronto, Ontario M5S 1A7, Canada}

\author{Asimpunya Mitra}
\affiliation{Department of Physics, University of Toronto, Ontario M5S 1A7, Canada}

\author{Félix Desrochers}
\affiliation{Department of Physics, University of Toronto, Ontario M5S 1A7, Canada}

\author{Emily Z. Zhang}
\affiliation{Department of Physics, University of Toronto, Ontario M5S 1A7, Canada}

\author{Lubuna Shafeek}
\affiliation{Institute of Science and Technology Austria, 3400 Klosterneuburg, Austria}

\author{Austin Ferrenti}
\affiliation{Institute for Quantum Matter and William H. Miller III Department of Physics and Astronomy,
Johns Hopkins University, Baltimore MD 21218, USA}
\affiliation{Department of Chemistry, The Johns Hopkins University, Baltimore, MD 21218, USA}

\author{Tyrel M. McQueen}
\affiliation{Institute for Quantum Matter and William H. Miller III Department of Physics and Astronomy,
Johns Hopkins University, Baltimore MD 21218, USA}
\affiliation{Department of Chemistry, The Johns Hopkins University, Baltimore, MD 21218, USA}

\author{Arkady Shekhter}
\affiliation{Los Alamos National Laboratory, Los Alamos New Mexico 87545, USA}

\author{Zolt\'an K\"oll\"o}
\affiliation{Institute of Science and Technology Austria, 3400 Klosterneuburg, Austria}

\author{Yong Baek Kim}
\affiliation{Department of Physics, University of Toronto, Ontario M5S 1A7, Canada}

\author{B.~J.~Ramshaw}
\affiliation{Laboratory of Atomic and Solid State Physics, Cornell University, Ithaca, NY, USA}
\affiliation{Canadian Institute for Advanced Research, Toronto, Ontario, Canada}

\author{K.~A.~Modic}
\email{kimberly.modic@ist.ac.at}
\affiliation{Institute of Science and Technology Austria, 3400 Klosterneuburg, Austria}

\date{\today}

\begin{abstract}
\bf
Early efforts to realize exotic quantum ground states in frustrated magnets focused on frustration arising from the lattice geometry alone. Attention has shifted to bond-dependent anisotropic interactions, as well as further-neighbor interactions, on non-geometrically-frustrated lattices due to their greater versatility. The honeycomb magnet \bcao recently emerged as a candidate host for both bond-dependent (e.g. Kitaev) and third-neighbor ($J_3$) interactions, and has become a model experimental system due to its relatively low levels of disorder. Understanding the relative importance of different exchange interactions holds the key to achieving novel ground states, such as quantum spin liquids. Here, we use the magnetotropic susceptibility to map out the intermediate and high-field phase diagram of \bcao as a function of the out-of-plane magnetic field direction at $T = 1.6$ K. We show that the experimental data are qualitatively consistent with classical Monte Carlo results of the \xxz model with small Kitaev and off-diagonal exchange couplings included. However, the calculated critical fields are systematically larger than the experimental values. Infinite-DMRG computations on the quantum model reveal that quantum corrections from a nearby ferromagnetic state are likely responsible for the suppressed critical fields. Together, our experiment and theory analyses demonstrate that, while quantum fluctuations play an important role in determining the phase diagram, most of the physics of \bcao can be understood in terms of the classical dynamics of long-range ordered states, leaving little room for the possibility of a quantum spin liquid.
\end{abstract}

\maketitle

\section*{Introduction}
\label{intro}

The spin-orbit entangled $J_{\rm eff}=1/2$ local moments in Co-based frustrated magnets offer novel routes to achieve exotic quantum ground states, such as a quantum spin liquid (QSL). One example is \bcao, with a low-disorder honeycomb lattice and an ordering temperature that is suppressed well below the exchange energy scale \cite{Zhong_2020}. The lattice geometry and spin-orbit coupling allow for bond-dependent interactions, such as the Kitaev interaction, and third nearest-neighbour exchange interactions, both of which lead to significant magnetic frustration. Recently explored in the context of Kitaev magnetism \cite{Kitaev_2006, Takagi_2019, Liu_2018, Sano_2018, Liu_2020}, \bcao is likely better described by the \xxz model where the XXZ anisotropy comes from the local trigonal distortion of the oxygen octahedra \cite{Liu_2023, Jiang_2023, Maksimov_2023, Bose_2023, Das_2021, Maksimov_2022, Winter_2022}. In addition, the inclusion of small anisotropic interactions is required to explain the gap in the magnon spectrum \cite{Halloran_2022}. The rich field-temperature phase diagram of \bcao makes it an ideal test of our ability to model frustrated spin-1/2 systems \cite{Regnault_1979, Regnault_2018, Zhong_2020}. Furthermore, terahertz measurements recently reported the emergence of an excitation continuum with the application of a magnetic field perpendicular to the basal plane~\cite{Zhang_2023}. This continuum was interpreted as a signature of a field-induced QSL, and calls for a critical examination of the potential stabilization of exotic ground states in \bcao through the application of an out-of-plane magnetic field.

Classical Monte Carlo and molecular dynamics simulations were previously used to determine a minimal spin model by fitting neutron scattering and thermodynamic data \cite{Halloran_2022}. The resulting \xxz spin Hamiltonian provides an excellent description of the dynamical spin structure factor in zero magnetic fields and in the high-field polarized state. The classical model further captures the incommensurate ordering wave vector $|\mathbf{q}_c|=0.27$ along the $\Gamma\to M$ direction of the zero-field ground state, and the intermediate commensurate field-induced phase with $|\mathbf{q}_c|=1/3$ observed experimentally for an in-plane field between $B_{c1}=0.33$~T and $B_{c2}=0.55$~T~\cite{Zhong_2020, Regnault_1979, Regnault_1990, Regnault_2018, Halloran_2022}. This intermediate phase is stabilized in the classical model by weak bond-dependent interactions \cite{Halloran_2022}. Despite this success, the classical model’s ability to capture the system’s behaviour upon the application of a field with an out-of-plane component has yet to be tested. Indeed, if the field-induced QSL suggested in Ref.~\cite{Zhang_2023} is realized, significant qualitative discrepancies between the classical model and experimental observations are expected as the QSL cannot be realized in the classical model. This discourse calls for sensitive measurements of the magnetic response as a function of the external magnetic field's direction. Detailed angular information can further provide invaluable information in determining the correct values of the small anisotropic terms, which play a crucial role when a large number of states are in close energetic competition, such as in \bcao. Moreover, these subleading interactions may be important to understand the short-range quasi-collinear (double-zig-zag-like) correlations in the incommensurate ordered state observed in spherical neutron polarimetry experiments~\cite{Regnault_2018}. Currently, only an incommensurate spiral and the commensurate double zig-zag state with $|\mathbf{q}_c|=1/4$ are realized in relevant classical and quantum models, respectively~\cite{Halloran_2022,Maksimov_2022b, Jiang_2023}.

In this work, we measure the magnetotropic susceptibility -- the second derivative of the free energy with respect to magnetic field angle -- of \bcao at $T = 1.6$ K \cite{Modic_2018, Shekhter_2023}. Through precise angle-dependent measurements, we map out the critical fields for both the incommensurate to commensurate transition, $B_{c1}$, and the commensurate to field-polarized transition, $B_{c2}$ (\autoref{fig:known diagram}). While the Monte Carlo results of the classical \xxz model capture the qualitative features of the field-angle phase diagram, and the ratio $B_{c2}/B_{c1}$, for a wide range of angles, they overestimate the overall critical field scale by a factor of roughly 1.6. Using infinite density matrix renormalization group (iDMRG)~\cite{White_1992, White_1993, Schollwock_2005, McCulloch_2008, Hauschild_2018}, we obtain an incommensurate ground state that is stable over a much-reduced parameter regime than predicted classically. Hence, quantum corrections to the ground state energies are likely responsible for the overestimation of the critical fields in the classical calculations. 


\section*{Experiment}

We use resonant torsion magnetometry \cite{Modic_2018} to measure the magnetotropic susceptibility of \bcao \cite{Shekhter_2023}. Samples were prepared as reported in \citet{Ferrenti_2023} (see SI \ref{section:crystal} for additional details). Small single crystals are cut with focused-ion beam (FIB) lithography and aligned with electron back-scattered diffraction (EBSD) (see \autoref{fig:angle data} and the SI). These samples are welded onto a silicon micro-cantilever with FIB-deposited platinum. Measurements on FIB'ed samples were cross-checked against measurements on larger, non-FIB'ed, single crystals. Measurement on FIB'ed crystals is advantageous since their smaller size reduces the signal and thus allows us to access the entire field-angle phase diagram of a system with large magnetic anisotropy. 

\begin{figure}
\begin{center}
\includegraphics[width=.70\linewidth, trim=0cm 0cm 0cm 0cm, clip=true]{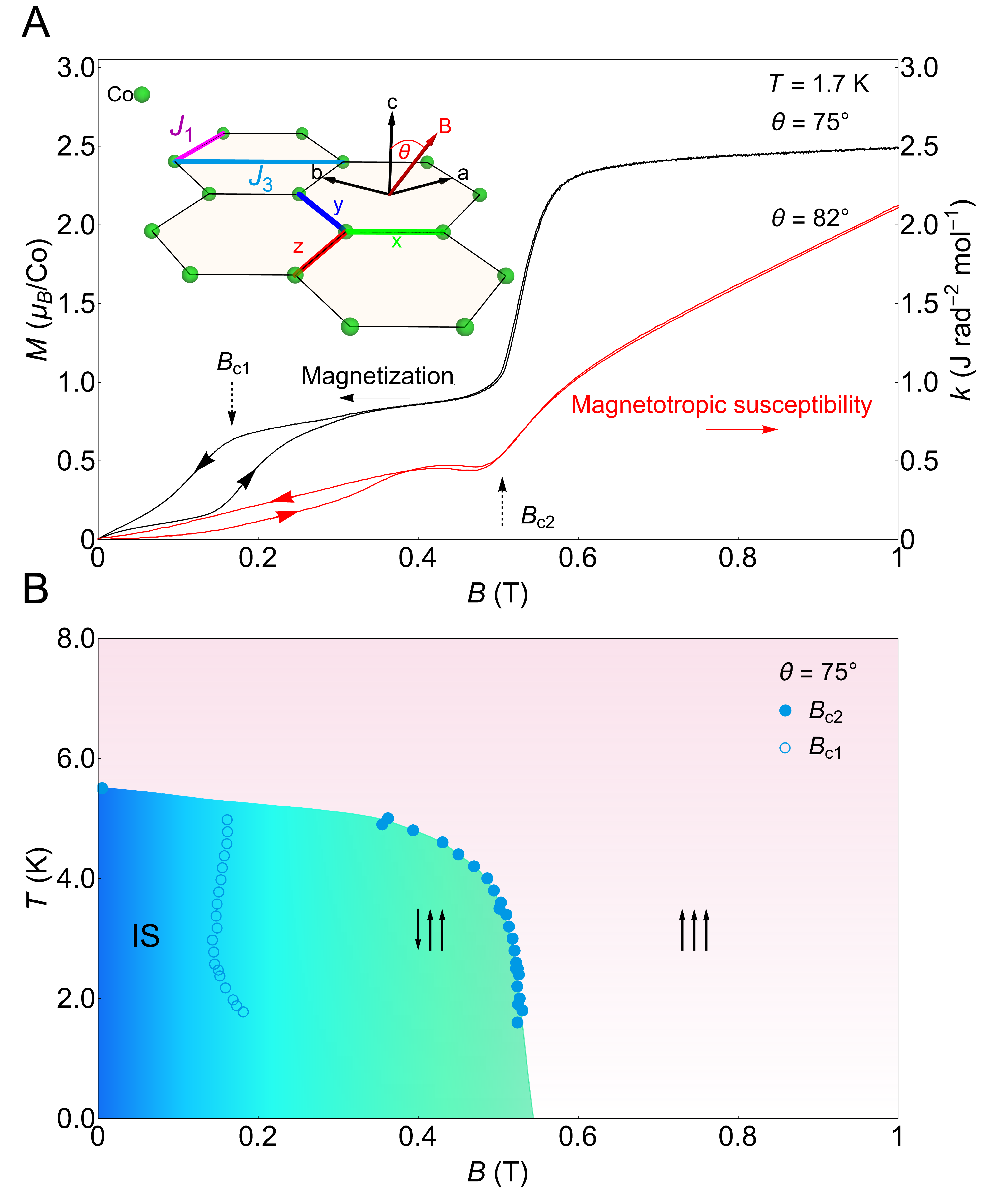}
\end{center}
\caption[Crystal]{\scriptsize \textbf{Established in-plane phase diagram of \bcao.} A) Magnetization (left axis) and magnetotropic susceptibility (right axis) measured for a nearly in-plane magnetic field ($\theta = 75^\circ$) to show how features correlate across different measurements techniques. Hysteresis is observed around the incommensurate to commensurate transition, \bco, in both curves. The sharp increase to saturated magnetization at 0.5 T corresponds to a minimum in the magnetotropic susceptibility, with the slight mismatch in field position due to the small angular difference. The inset shows the honeycomb network of cobalt ions highlighting the first- and third-neighbor interactions, $J_1$ and $J_3$, respectively. The crystallographic axes $a,b,c$ are shown with respect to the honeycomb bonds and $\theta$ is defined as the angle between the magnetic field direction and the $c$-axis. B) The temperature-magnetic field phase diagram of \bcao for $\theta = 75^\circ$ constructed from magnetization and magnetotropic susceptibility data (IS = incommensurate state).}
\label{fig:known diagram}
\end{figure}

We measure the resonant frequency of the sample-cantilever system at its fundamental mechanical mode (\autoref{fig:angle data}). Changing the magnetic field and field angle leads to shifts in the system's resonant frequency due to the magnetic anisotropy in the sample. These shifts are proportional to the second derivative of the free energy as a function of magnetic field angle (\textit{i.e.} the magnetotropic susceptibility \cite{Modic_2018, Shekhter_2023}). In the linear response regime (\textit{i.e.} when the magnetization grows linearly with magnetic field), the magnetotropic susceptibility is equal to $k = \mu_o(\chi_{ii}-\chi_{jj}) H^2 \cos 2 \theta$. Here, $i$ and $j$  reflect the principal axes of the magnetic susceptibility tensor that lie in the plane of vibration, and $\theta$ defines the angle between the crystallographic $c$-direction and the magnetic field (see \autoref{fig:known diagram}). 

\section*{Results}
\subsection{Establishing the correspondence between the magnetization and the magnetotropic susceptibility}

The distinct regions of the temperature-magnetic field phase diagram of \bcao have been extensively studied under an in-plane magnetic field \cite{Regnault_1979, Regnault_2018, Zhong_2020, Shi_2021, Halloran_2022, Zhang_2023} and are in agreement with our data for a largely in-plane field component in \autoref{fig:known diagram}B. \bcao enters an incommensurate antiferromagnetic (AFM) phase at $T_\text{N} = 5.5$ K at zero magnetic field. Under an applied magnetic field, the incommensurate phase gives way to a commensurate phase with an $\uparrow \uparrow \downarrow$ structure, consistent with the magnetization reaching 1/3 of the full saturated moment in the range from  0.2 T $\lesssim$ $B$ $\lesssim$ 0.5 T (left axis of \autoref{fig:known diagram}A) \cite{Zhong_2020, Shi_2021, Halloran_2022}. At 1.5 K and above 0.5 T, an in-plane magnetic field polarizes the spins and saturates the magnetization at 2.5 $\mu_B$ per Co$^{2+}$ --- consistent with the calculated $g$-factor \cite{Zhong_2020}.  

\begin{figure}[htbp]
\centering
\includegraphics[width=1\textwidth, trim=0cm 0cm 0cm 0cm, clip=true]{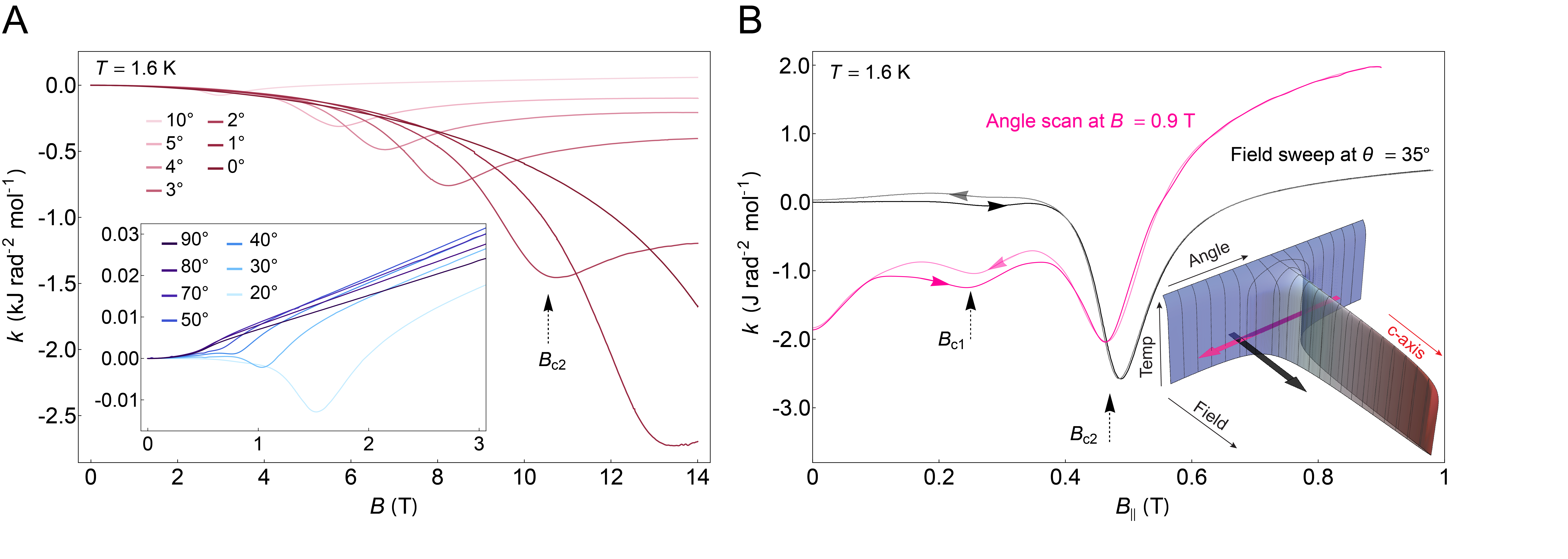}
\rule{27em}{0.5pt}
\caption[Crystal]{ \textbf{The out-of-plane magnetotropic susceptibility.} \scriptsize A) The magnetotropic susceptibility as a function of magnetic field applied along several directions near the $c$-axis. \bct is highly anisotropic for magnetic fields near the $c$-axis. The maximum \bct observed is $\sim13.5$ T. Fits to the position of the minima as a function of angle are used to extract a $c$-axis critical field of 15 T.  The inset shows similar data for field angles approaching the in-plane direction. B) The magnetotropic susceptibility measured as a function of the magnetic field at fixed angle (black curve) and as a function of angle in a fixed magnetic field (pink curve). We define the in-plane component of field as $B_{||} \equiv B \sin\theta$. \bco and \bct appear as minima at 0.25 T and 0.5 T, respectively, in both data sets when plotted versus $B_{||}$. The inset schematically shows how the AFM phase boundary is traversed for both measurements.  }
\label{fig:versus B}
\end{figure}

The step in the magnetization that marks the transition into the field-polarized phase correlates with a dip in the magnetotropic susceptibility (right axis of \autoref{fig:known diagram}A). This feature is expected because $k$ is proportional to the anisotropic susceptibility, which itself is a field-derivative of the magnetization. We identify this feature as \bct. The amplitude of the change in $k$ at \bct goes to zero along the high symmetry directions of the phase boundary (i.e. where $\partial T_\text{N} / \partial \theta  \rightarrow 0$, see the inset of \autoref{fig:versus B}B). We therefore show the magnetotropic data for $\theta = 75^\circ$ in order to highlight \bct. From field sweeps of $k$ (black curve in \autoref{fig:versus B}B), we also identify a region of hysteresis around a change in slope at \bco that corresponds to the transition between the incommensurate and commensurate phases. \bco, however, is much sharper when $k$ is measured as a function of field angle (pink curve in \autoref{fig:versus B}B). We therefore use the angle-sweep data to identify \bco and \bct in constructing the field-angle phase diagram of \bcao.

\subsection{Mapping the phase boundaries with angle}
We now identify the magnetic phase transitions as the field is rotated from the in-plane direction ($\theta = 90^\circ$) towards the $c$-axis ($\theta = 0^\circ$). The magnetic easy-axes lie in the honeycomb plane. Thus, \bct is pushed to higher magnetic fields for larger out-of-plane field components (\autoref{fig:versus B}A). \autoref{fig:versus B}B shows two curves that cut through the schematic phase boundary as a function of both magnetic field orientation and magnetic field strength. Since the phase boundaries in \bcao are predominantly sensitive to the in-plane field component, we plot the data as a function of $B\sin\theta$ and find that both magnetic phase transitions scale approximately as $B_c(\theta) = B_c(90^\circ)/|\sin\theta|$.

\begin{figure}[htbp]
\centering
\includegraphics[width=0.8\linewidth, trim=0cm 0cm 0cm 0cm, clip=true]{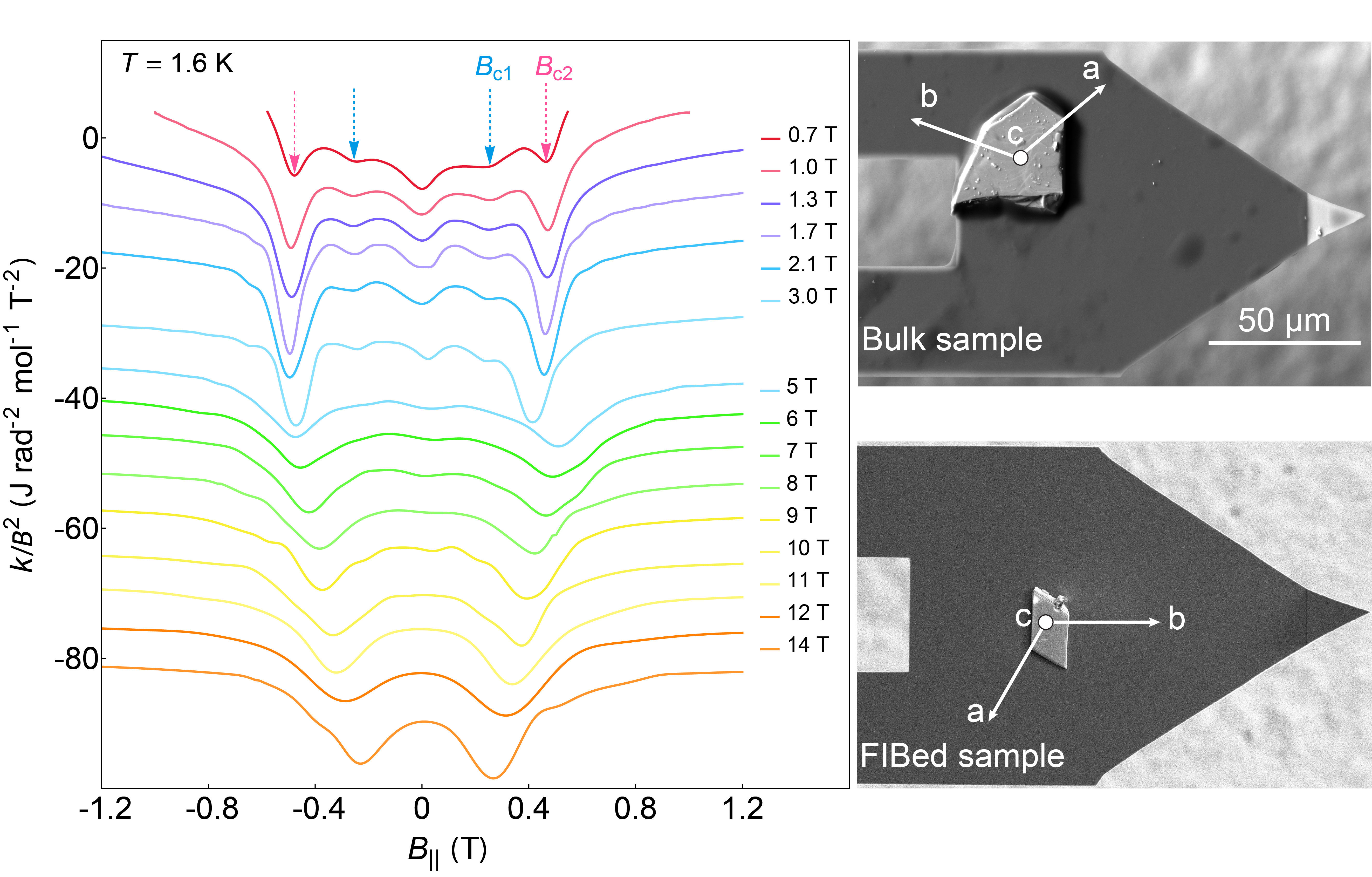}
\rule{27em}{0.5pt}
\caption[Crystal]{\scriptsize \textbf{Critical fields of \bcao at $\mathbf{ \textit{T} = 1.6}$ K.} The magnetotropic susceptibility normalized by $B^2$, plotted versus the in-plane magnetic field $B_{||}\equiv B \sin \theta$, and offset for clarity. The large change in $k$ near \bct for fields close to the $c$-axis limits our ability to track the resonant frequency at high fields. Therefore, to reduce the signal to a measurable range across all field angles, we measured a FIB'ed crystal with a volume $\sim$50 times smaller than the bulk sample for $B \geq 5$ T. SEM images of the two measured samples are shown next to the corresponding data.  The lower-field data ($\leq 3$ T) has been multiplied by 3.5 for clarity on this scale (see SI for details on unit conversions and scaling factors between cantilevers and crystals). }
\label{fig:angle data}
\end{figure}

We find that \bco is more pronounced in measurements of $k$ as a function of magnetic field orientation, rather than field strength (\autoref{fig:versus B}B). We therefore use angle-dependent measurements to map both critical fields across the entire phase diagram. \autoref{fig:angle data} shows the magnetotropic susceptibility at $T= 1.6$ K on two different samples: low-field measurements (up to 3 T) on the larger bulk crystal provide higher resolution of the intermediate field-induced phase, and the smaller FIB'ed crystal is used to track the large shift in resonant frequency through \bct to higher magnetic fields (from $B = 5$ to 14 T). Both  critical fields are plotted as a function of field angle in \autoref{fig:phase diagram}.




\autoref{fig:angle data} illustrates that the critical fields are almost entirely determined by the in-plane component of the magnetic field. Only when the field is very close to the $c$-axis is the divergence of the critical field cut off to a finite value (\autoref{fig:phase diagram}). This behavior (for both \bco and \bct) is captured by
\begin{equation}
	B_{\text{c}} = \frac{B^0_{\text{c}}}{\sqrt{\sin^2\theta+\frac{1}{\gamma^2}\cos^2\theta}},
	\label{eq:Bc_fit}
\end{equation}
where $B^0_{\text{c}} \equiv B_{\text{c}}(\theta =90^\circ)$ is the in-plane critical field, and $\gamma$ is the anisotropy parameter defined such that $B^0_{\text{c}}\gamma$ is the out-of-plane critical field. We find $\gamma = 31$ for \bct, which means that the correction to $\text{sin}\theta$ scaling is not resolvable in the data until $\theta \lesssim 2^\circ$ (\autoref{fig:ratios}). This is significantly more anisotropic than RuCl$_3$, where $\gamma$ is less than 10 \cite{Modic_2020, Zhou_2023}. \bco cannot be tracked to low enough angle to obtain a reliable fit for $\gamma$. With the field-angle phase diagram determined, we now investigate whether the \xxz model can provide a good description of both its qualitative and quantitative features.

\subsection{Monte Carlo simulations of the classical model}

Using classical Monte Carlo methods, we compute the magnetotropic susceptibility for various field strengths and angles. We use an \xxz model with small bond-dependent anisotropies presented by \citet{Halloran_2022} with the addition of a Zeeman term coupling spins to the magnetic field. We write this Hamiltonian as $\mathcal{H}=\mathcal{H}_\mathrm{XXZ}+\mathcal{H}_\mathrm{anisotropy}+\mathcal{H}_\mathrm{Zeeman}$.

\begin{equation}
    \begin{split}
\mathcal{H}_\mathrm{XXZ}&=\sum_{n=1,3}\sum_{\langle i, j\rangle_n}\left(J_{xy}^{(n)}\left(S_i^xS_j^x+S_i^yS_j^y\right)+J_z^{(n)}S_i^zS_j^z\right)\\
\mathcal{H}_\mathrm{anisotropy}&=\sum_{\langle i, j \rangle \in \lambda}\Bigl(D_\lambda\left(S_i^xS_j^x-S_i^yS_j^y\right)+E_\lambda\left(S_i^xS_j^y+S_i^yS_j^x\right)\Bigr)\\
    \mathcal{H}_\mathrm{Zeeman}&=-\mu_\mathrm{B}B\sum_i\Bigl(g_{ab}\sin(\theta)\left(S_i^x\cos(\phi)+S_i^y\sin(\phi)\right)+g_c\cos(\theta)S_i^z\Bigr) 
    \end{split}\label{eq:Hamiltonian}
\end{equation}
where $\lambda=\{x, y, z\}$ are labels for the three different first nearest-neighbour bond directions. The details of the bond-dependent couplings, $D_\lambda$ and $E_\lambda$, are presented in the SI \ref{section:Model Parameters}. The exchange parameters used in the current work are slightly different from those in \citet{Halloran_2022}. The refined parameter set is presented in the SI \ref{section:Model Parameters} and provides a better agreement with the magnetization data. Because all observables are largely independent of the angle $\phi$ (measured away from the $a$-axis), both in experiments and in our simulations \cite{Halloran_2022}, we choose to explore $\phi=0$ for simplicity. The calculations were performed at $ T = 1.6$ K to match the experimental temperature. 

\begin{figure}[htbp]
\centering
\includegraphics[width=0.7\linewidth, trim=0cm 0cm 0cm 0cm, clip=true]{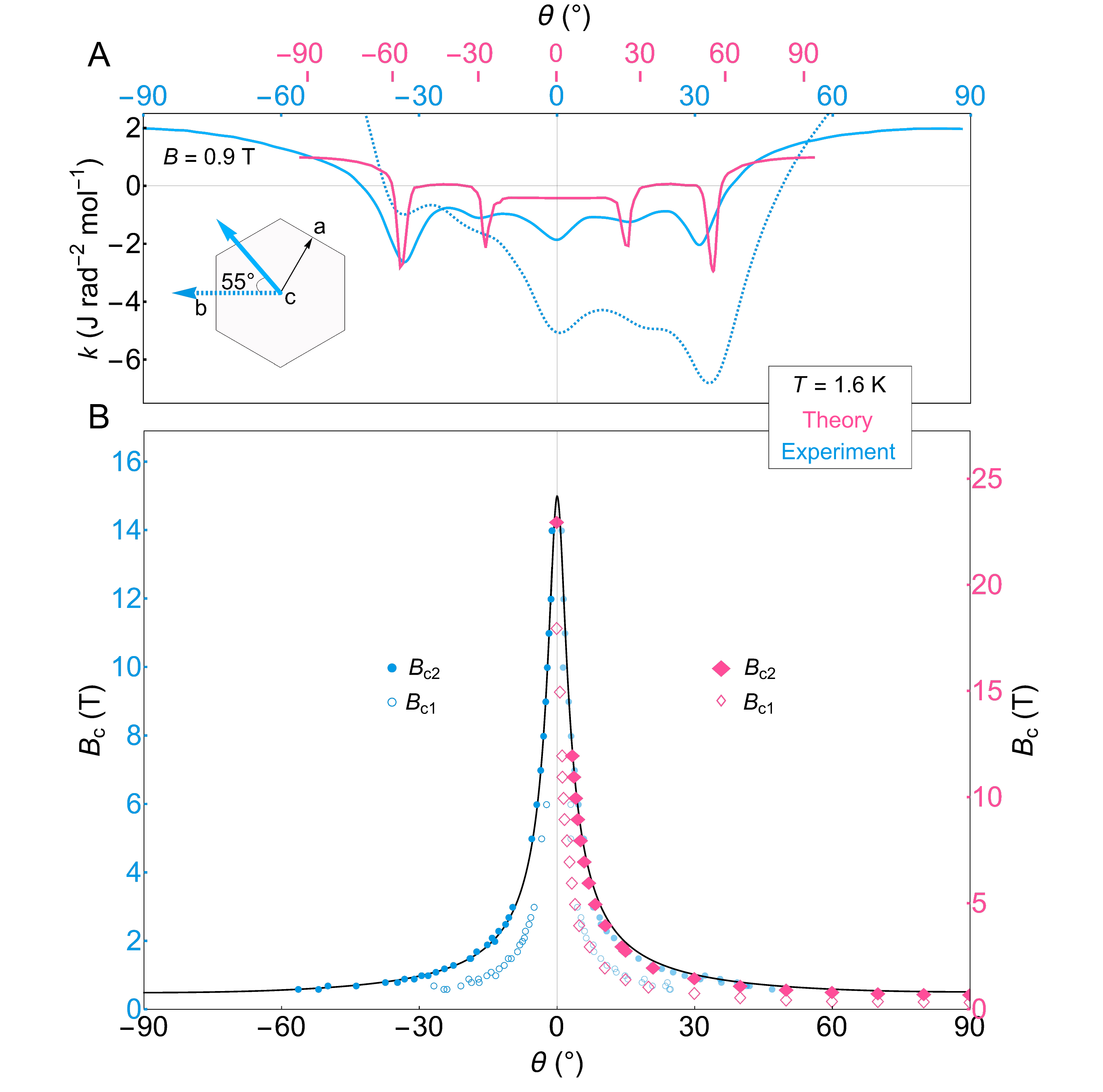}
\rule{27em}{0.5pt}
\caption[Crystal]{\textbf{Comparison between experimental and theoretical results.} \scriptsize A) The magnetotropic susceptibility of \bcao at $ T = 1.6$ K and $B = 0.9$ T for rotation of magnetic field in two different crystal planes: the dashed blue line corresponds to rotation from the $c$-axis ($\theta = 0^\circ$) towards the $b$-axis ($\theta = 90^\circ$), and the solid blue line shows rotation from the $c$-axis into the $ab$-plane at an angle $\sim 55^\circ$ from the $b$-axis. The colored arrows on the honeycomb (inset) illustrate the two planes of rotation. The solid blue data has been scaled by 3.5. The angle in the Monte Carlo results has been scaled by 1/1.6, the same scaling factor needed to match the experimental and theoretical critical fields (\autoref{fig:ratios}). B) The pink points show the transition fields from Monte Carlo results plotted versus field angle. The blue points show \bco and \bct selected from both angle sweep (\autoref{fig:angle data}) and field sweep data (\autoref{fig:versus B}). The black lines through the \bct data points are fits of the experimental (theoretical) data to \autoref{eq:Bc_fit} on the -$\theta$ (+$\theta$) side. The calculations of the critical fields qualitatively capture the angle dependence of both \bco and \bct, but the overall field scale is 1.6 times larger than observed.} 
\label{fig:phase diagram}
\end{figure}

We identify two minima in the magnetotropic susceptibility that correlate well with the transition to the 1/3 magnetization plateau and to the fully polarized state \autoref{fig:phase diagram}A. This is qualitatively similar to the experimental data. We further confirm these phases to be the incommensurate spiral ($|\mathbf{q}_c|\approx0.29$), the $|\mathbf{q}_c|=1/3$ phase, and the polarized paramagnet by examining the peaks in the static spin structure factor.

By tracking the two minima in the calculated magnetotropic susceptibility as a function of angle, we produce the phase diagram shown in \autoref{fig:phase diagram}B. The shapes of the resulting curves ($B_{c1}(\theta)$ and $B_{c2}(\theta)$) match extremely well with experiment. We also compute $\gamma$ for $B_{c2}$ and find $\gamma=30$. This is in agreement with the experimentally determined value of 31 above. We note that while the qualitative features of the calculated phase diagram are in good agreement with those of the experimental phase diagram, the overall calculated field scale is larger by more than 50\%. 


\begin{figure}[htbp]
\centering
\includegraphics[trim=0cm 0cm 0cm 0cm, clip=true]{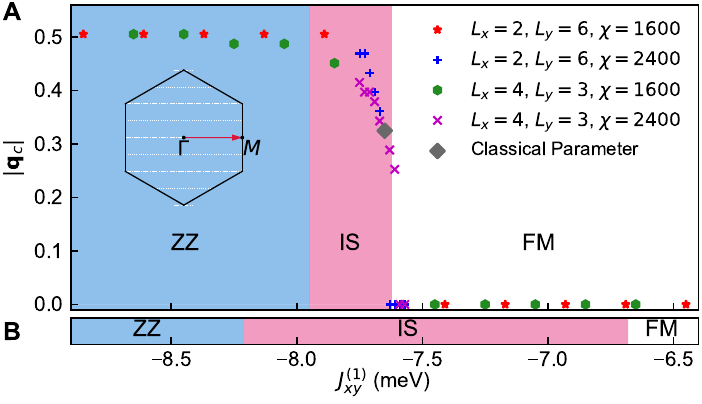}
\rule{27em}{0.5pt}
\caption[Crystal]{ \scriptsize \textbf{Quantum Phase diagram using iDMRG} (A) The position of the peak in the static spin structure factor $|\mathbf{q_c}|$ is used to identify the various phases: ferromagnet (FM, $|\mathbf{q}_c|=0.0)$, zig-zag (ZZ, $|\mathbf{q}_c|=0.5)$, and incommensurate state (IS) with $|\mathbf{q}_c|$ at wavevectors between $\Gamma-M$ (along red arrow in inset). An XC cylindrical geometry was used in iDMRG with an MPS unit cell size $L_y$ along the circumference, $L_x$ along the infinite direction, and a bond dimension $\chi$. The BZ in the inset shows the momentum cuts along which the spin structure factor can be accessed in this cylindrical geometry, dotted lines for $L_y=6$, and dashed-dotted lines for both $L_y=3$ and $L_y=6$. We have used $J_z^{(1)}=-1.2$ meV, $J_z^{(3)}=-0.8$ meV and the constraint $J_{xy}^{(1)}+J_{xy}^{(3)}=-5.0$ meV. At the best-fit classical parameter of $J_{xy}^{(1)}=-7.65$ meV, the ground state (grey diamond) is in the IS phase, but is close to the phase boundary between the FM and the IS. (B) The classical phase boundaries obtained using variational single Q-ansatz and simulated annealing for the same parameters as in (A). 
The region of stability of the IS in the quantum phase diagram is reduced compared to (B) due to quantum fluctuations.
}
\label{fig:iDMRG}
\end{figure}

\subsection{iDMRG quantum phase diagram}

To investigate the quantitative discrepancy in the values of the critical fields between the experiment and Monte-Carlo simulations, we use iDMRG \cite{White_1992,White_1993,Schollwock_2005,McCulloch_2008} of the quantum model to study the nature of the ground state at zero field. We simulate the \xxz model without the $D_{\lambda}$ and $E_{\lambda}$ anisotropies (SI \ref{section:Model Parameters}). These small bond-dependent interactions are omitted to numerically leverage the continuous $U(1)$ symmetry of the $\mathcal{H}_{\mathrm{XXZ}}$ Hamiltonian. A similar model was recently investigated with DMRG~\cite{Jiang_2023}, variational Monte-Carlo~\cite{Bose_2023}, and pseudo-Fermion functional renormalization group~\cite{Watanabe_2022}. Unlike DMRG on a finite slab, iDMRG on an infinite cylinder can capture incommensurate orders with a continuously varying ordering wavevector, which is of particular relevance for the ground state of \bcao. 

Varying the ratio of first $J_{xy}^{(1)}$ and third-nearest-neighbour in-plane coupling $J_{xy}^{(3)}$, we find magnetically ordered phases that can be identified by tracking the position of the magnetic Bragg peak in the static spin structure factor. In the quantum phase diagram shown in \autoref{fig:iDMRG}A, we can clearly identify the ferromagnet $(|\mathbf{q}_c|=0.0)$, the zig-zag $(|\mathbf{q}_c|=0.5)$, and the incommensurate ordered phases with peaks at continuously varying points $|\mathbf{q}_c|$ along the $\Gamma-M$ direction (see inset). Compared with the classical phase boundaries in \autoref{fig:iDMRG}B, the parameter regime where the incommensurate order is stabilized is greatly reduced in the quantum model. In particular, we find that the quantum ground state using the exchange couplings for $\mathcal{H}_{\mathrm{XXZ}}$ used in this work is incommensurate (SI \ref{section:Model Parameters}), but in very close proximity to the phase boundary of the ferromagnetic state. In addition, if we use the set of exchange parameters (excluding small anisotropy terms) considered in a previous DMRG study \cite{Jiang_2023}, we obtain a ferromagnetic ground state. This is consistent with their result and our conclusion of the proximity to the ferromagnetic state. Note that the parameters in \citet{Jiang_2023} are slightly different from the original parameters in \citet{Halloran_2022}, as well as the refined set used in the current work. Detailed comparison between these results is discussed in the SI \ref{section:iDMRG}. The clear reduction of the incommensurate state's stability suggests that quantum fluctuations of the ferromagnetic state may suppress the critical fields of the nearby incommensurate state.

\section*{Discussion}

The classical simulations reproduce several features of the experimental phase diagram. For example, the calculated ratio  of \bct to \bco is $\sim$1.9, which agrees well with the experimentally-determined value of $\sim$1.9 (SI \ref{section:Magnetotropic}).  The classical simulations find that the anisotropy in \bct is $\gamma = 30$, again very close to the experimental value of $\gamma = 31$. Even the structure of the simulated magnetotropic susceptibility data is qualitatively consistent with the measured data, revealing two distinct peaks that correspond to phase transitions from the incommensurate phase to the $|\mathbf{q}_c|=1/3$ phase and then from the $|\mathbf{q}_c|=1/3$ phase to the polarized paramagnet. Further, the change in the magnitude of the calculated magnetotropic susceptibility at \bco and \bct is within a factor of two of the measured values (\autoref{fig:phase diagram}).  

The most obvious discrepancy between the data and the Monte Carlo simulations of the classical model is that the simulations overestimate the critical fields by roughly a factor of 1.6. The iDMRG results shed light on this issue. We find that iDMRG simulations using roughly the same best-fit exchange parameters obtained from the classical simulations find an incommensurate ordered state that is very close to the phase boundary with the nearby ferromagnetic state (see \autoref{fig:iDMRG}). The enhanced stability of the ferromagnetic state in the quantum model explains the decrease in the critical fields in the experiment as likely due to quantum zero-point fluctuations in the ordered phases. 



The level of qualitative agreement between experiments and Monte Carlo simulations of the classical model indicates that, even though quantum fluctuations are important in determining the quantitative position of the phase boundaries and corresponding ground states, most of the physics can be understood in terms of classical dynamics of long-range ordered states. Our measurement and analysis further suggest that the incommensurate state remains the ground state for an out-of-plane field up to more than $B = 10$ T. This observation is at odds with Ref.~\cite{Zhang_2023}, where a field-induced QSL was proposed to be stabilized for a magnetic field beyond $B = 4.5$ T. These considerations thus cast doubt on previous claims of the realization of QSL physics in \bcao for the explored parameter regime. Our detailed phase diagram (\autoref{fig:phase diagram}) is highly suggestive of the following interpretation: due to the gigantic easy-plane anisotropy of \bcao, the slight misalignment of the crystal with respect to magnetic field identified by \citet{Zhang_2023} causes the system to enter the intermediate $|\mathbf{q}_c|=1/3$ phase, rather than the field-polarized phase, for a field of approximately $B = 4$ T. Thus, the observed change in the terahertz spectrum at $B = 4$ T could correspond to a change in the magnon gap as the system transitions from the incommensurate to the intermediate phase at $B_{c1}$, rather than the appearance of a QSL at $B_{c2}$. This interpretation is supported by preliminary inelastic neutron scattering experiments~\cite{Halloran_2023}.

While many aspects of the \bcao phase diagram are captured by the \xxz model, the importance of other symmetry-allowed anisotropic spin interactions remains an open question. For example, one feature of the data that is not captured in the classical simulations is the asymmetry in the magnitude of $k$ when rotating from $\theta\xrightarrow{}-\theta$ (\autoref{fig:phase diagram}A). This asymmetry is not present in our MC calculations due to an underlying symmetry of the model Hamiltonian. Specifically, note that taking $\theta\xrightarrow{}-\theta$ is equivalent to taking $B_x\xrightarrow{}-B_x$ and $B_y\xrightarrow{}-B_y$ while leaving $B_z$ unchanged. The interacting part of our model Hamiltonian also has such an in-plane inversion symmetry under $(S_x, S_y)\xrightarrow{}(-S_x, -S_y)$. When taking $\theta\xrightarrow{}-\theta$, $S_x$ can be re-labeled as $-S_x$ and likewise for $S_y$, leaving the partition function (and therefore the free energy and the magnetotropic susceptibility) unchanged. The asymmetric measurements of $k$ reveal the need for additional terms in the model Hamiltonian which couple the in-plane spin components ($S_x$ and $S_y$) to $S_z$.

While we expect all of the anisotropic interactions to be much smaller than the main XXZ exchange parameters, the determination of these small interactions would provide the accurate estimation of the sub-leading bond-dependent interactions such as the Kitaev and $\Gamma$ interactions in the local spin coordinate. This would help us understand potential avenues to enhance such bond-dependent interactions.

\newpage
\setcounter{secnumdepth}{3}
\section*{Methods}
\renewcommand{\thesubsection}{\textnormal{\Roman{subsection}}}

\subsection{Crystal synthesis}
\label{section:crystal}
Polycrystalline samples of \bcao were produced by the heating of well-ground mixtures of BaCO$_3$ (Strem Chemicals, 99.9\%), Co$_3$O$_4$ (NOAH Technologies, 99.5\%, 325 mesh), and NH$_4$H$_2$AsO$_4$ (Alfa Aesar, 98\%) in a 2.9 : 2 : 6 molar ratio. The mixture was placed in an uncovered alumina crucible in air, heated to 305°C, held for 12 hours, heated to 875°C, held for 48 hours, cooled to room temperature, and finally heated to 925°C for 12 hours before again cooling to room temperature. All ramp rates were set at 100°C per hour. 

While \bcao melts incongruently under most atmospheric conditions, we found that single crystals could be grown via a modified Bridgeman technique. Polycrystalline samples were packed into a cylindrical alumina crucible with a pointed tip, which was then sealed under vacuum in a 12mm ID, 16mm OD quartz tube. The tube was then suspended in a vertical tube furnace, just above a hollowed block of firebrick.  The sample was heated to 1175°C at a rate of 100°C per hour and held for 12 hours without translation to ensure a homogeneous melt. The tube was then translated downward through the firebrick at a nominal rate of 3.0 mm per hour until the entire crucible had passed into the cooler zone. Furnace heating was then turned off and the sample was allowed to cool naturally within the firebrick. The static vacuum atmosphere was found to be necessary to achieve congruent melting in this system. The resulting mass of crystals was then pried from the crucible and mechanically separated along the cleavage planes.

\subsection{Vibrating sample magnetometry (VSM) data}
\label{section:VSM}

\subsubsection{Nearly in-plane magnetization}
\begin{figure}[htbp]
	\centering
	\includegraphics[width=.55\linewidth, trim=0cm 0cm 0cm 0cm, clip=true] {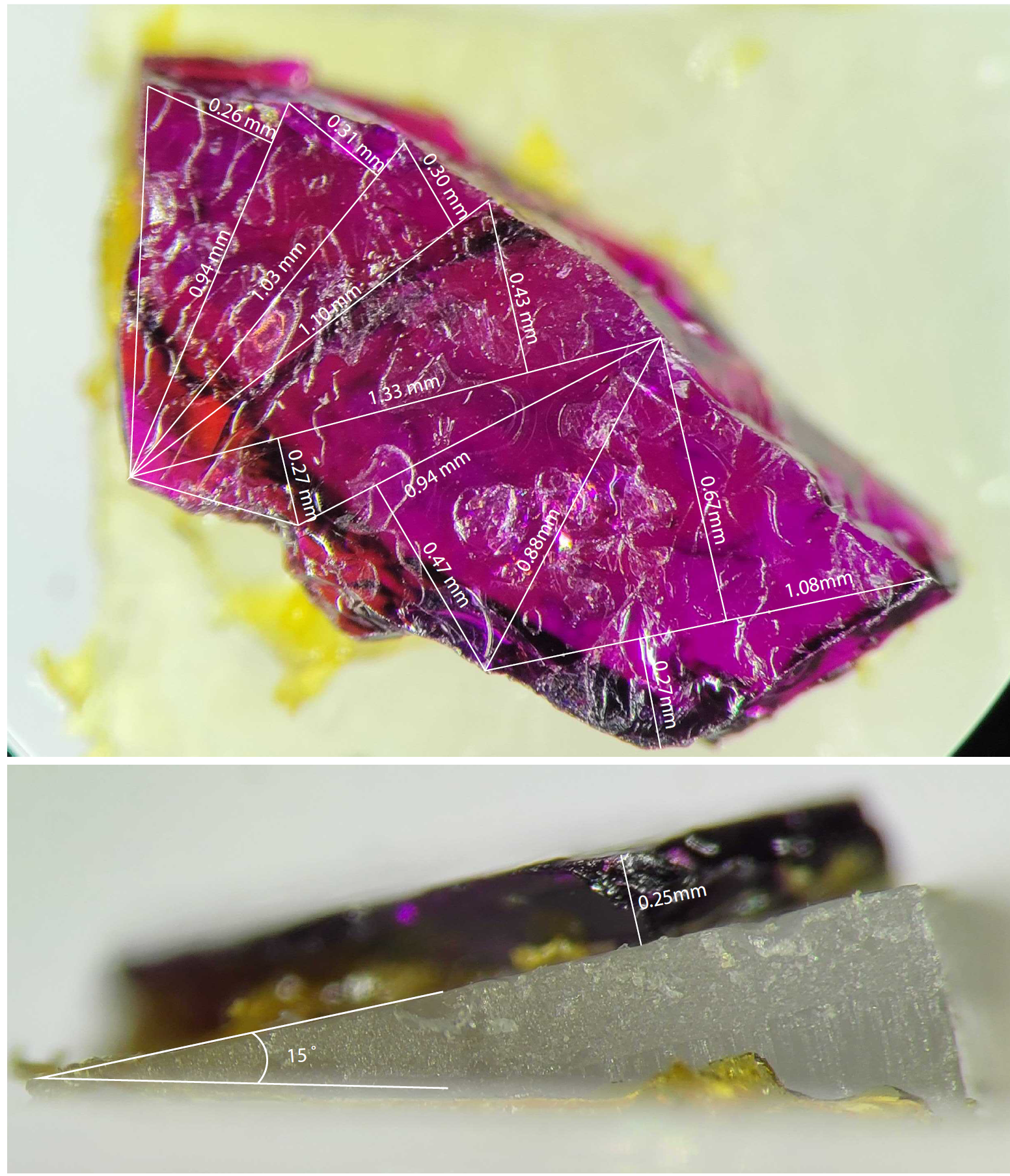}
	\rule{27em}{0.5pt}
	\caption[Crystal]{ \scriptsize \textbf{} A) Single crystal of \bcao B) Side view of the crystal in panel A mounted on a teflon wedge with a 15$^\circ$ angle using GE varnish. VSM measurements were performed on this sample with magnetic field applied at an angle $15^\circ$ from the honeycomb plane towards the $c$-axis.}
		\label{fig:CrystalPic}
	\end{figure}

\begin{figure}[htbp]
	\centering
	\includegraphics[width=0.75\linewidth, trim=0cm 0cm 0cm 0cm, clip=true]{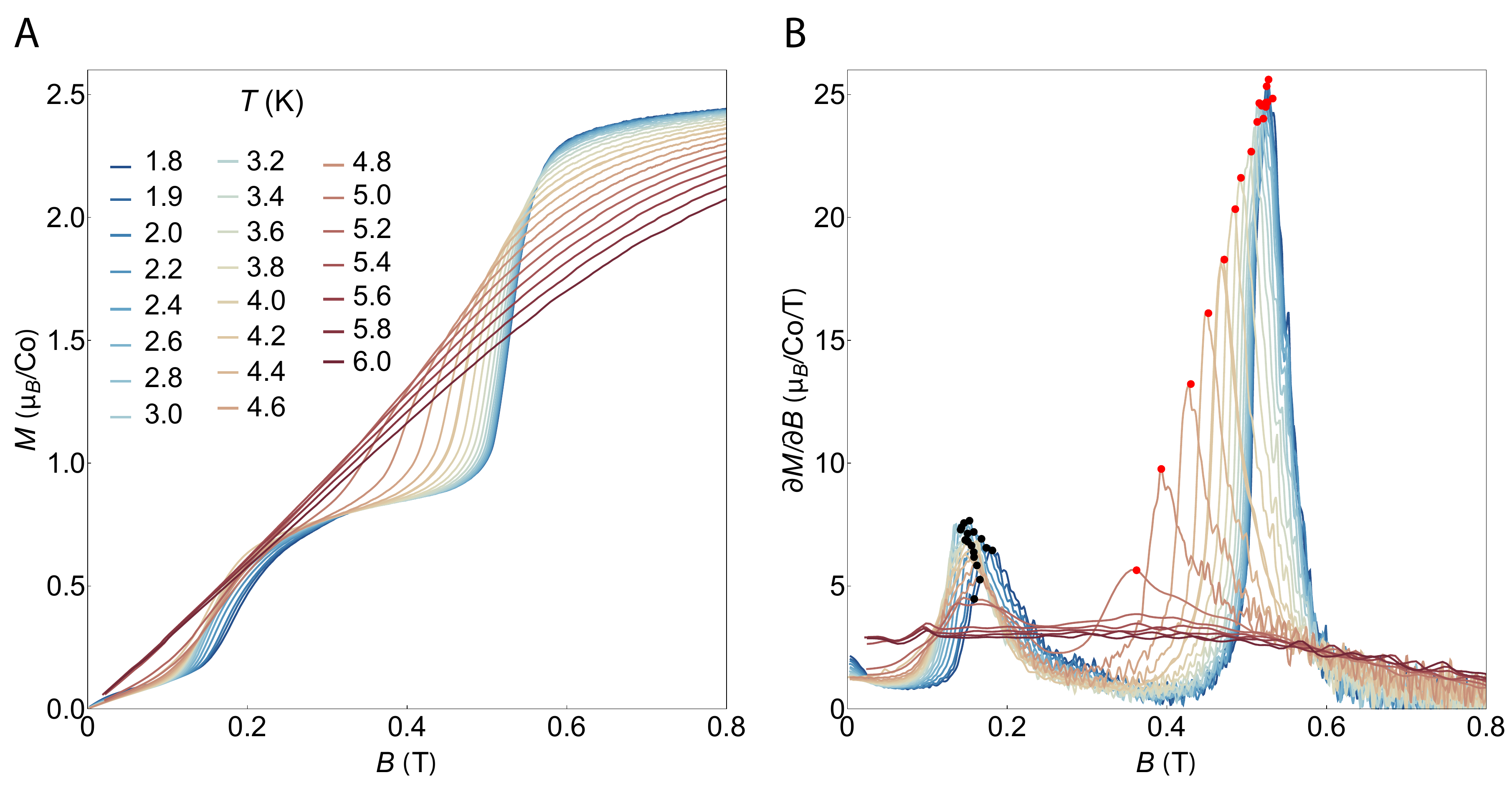}
	\rule{27em}{0.5pt}
	\caption[Crystal]{ \scriptsize \textbf{} A) Temperature-dependent magnetization of \bcao for field applied nearly in-plane ($\theta = 75^\circ$). B) The field-derivatives of the magnetization shown in panel A used to identify the positions of the critical fields, \bco and \bct.}
	\label{fig:NearlyInPlane}
	\end{figure}

Using vibrating sample magnetometry (VSM), we measured the magnetization of \bcao at several temperatures with the magnetic field applied at $75^\circ$ from the $c$-axis (\autoref{fig:NearlyInPlane}) in order to compare with magnetotropic measurements performed at the same field angle. A single crystal of \bcao with an approximate volume of 0.41 mm$^3$  (\autoref{fig:CrystalPic}A) contains approximately 4.8 $\times$ 10$^{18}$ cobalt atoms. We used the VSM option of a Quantum Design Dynacool Physical Property Measurement System (PPMS) with a 14 T magnet. The measured magnetization is read in emu units (1 = 1.08 $\times$ 10$^{20}$ $\mu_B$), which was divided by the estimated number of cobalt above to obtain the magnitude displayed in \autoref{fig:NearlyInPlane}A.

We find that the magnetization as a function of a nearly in-plane magnetic field is consistent with previous studies done with magnetic field applied $\textit{in}$ the honeycomb plane. As the lowest temperatures of $T = 1.8$ K, the magnetization saturates at $B = 0.5$ T with a magnetic moment of $\sim$2.5 $\mu_B$ per cobalt (\autoref{fig:NearlyInPlane}A). At low temperatures, there is also a clear transition into the $M = 1/3$ plateau in the field range from roughly 0.2 to 0.5 T. As temperature increases, the saturation field also increases. The field-derivatives of the magnetization curves in panel A are shown in panel B of \autoref{fig:NearlyInPlane}. The positions of \bco and \bct used to construct the phase diagram in (\autoref{fig:known diagram}B) are indicated with black and red points, respectively.

\subsubsection{Out of plane magnetization of \bcao}
We measured the magnetization of \bcao for magnetic field applied along the $c$-axis. Using the Photonic Science Laue X-ray Detector, we identified the $c$-axis of the sample and aligned it with the external magnetic field with $<$0.15$^\circ$ of uncertainty.
\begin{figure}[h]
    \centering
    \includegraphics[width=.65\linewidth, trim=0cm 0cm 0cm 0cm, clip=true]{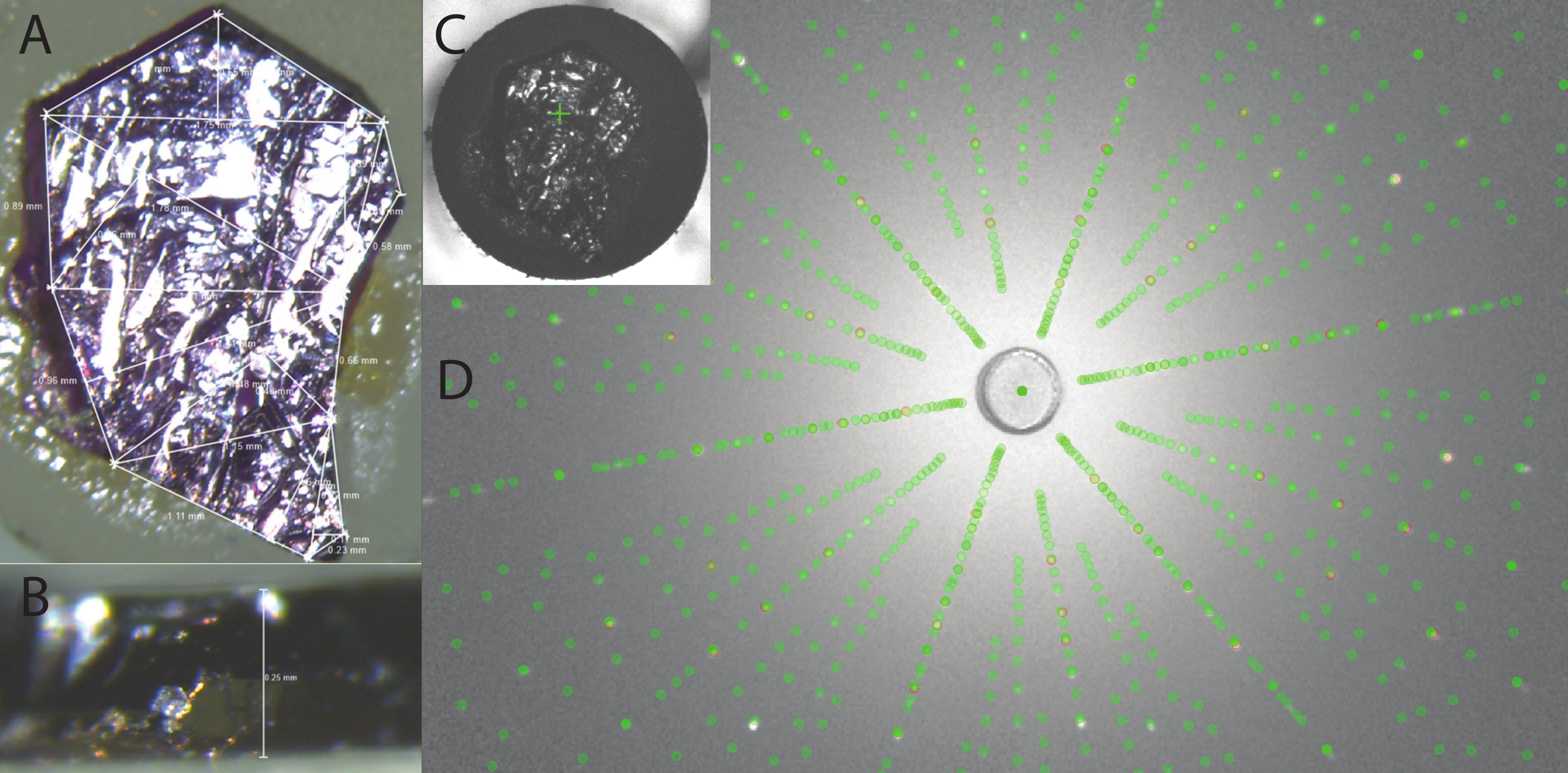}
    \rule{27em}{0.5pt}
    \caption{ \scriptsize \textbf{} A) Large single crystal of \bcao used for $c$-axis VSM measurements. B) The side view of the crystal in panel A mounted on a cylinder. The thickness of the crystal is. C) Optical image of the top of the sample mounted on a cylinder for Laue diffraction. D) The white points show the measured intensity of the diffraction pattern of \bcao overlaid with the simulated $c$-axis spectra in green.}
    \label{fig:Laue}
\end{figure}
The $c$-axis magnetization measurements were performed on the sample in Figure \ref{fig:Laue}, which has a volume of 0.87 mm$^3$ that consists of approximately 1 $\times$ 10$^{19}$ cobalt atoms. The measured $c$-axis magnetization in units of $\mu_B$ per cobalt is shown in \autoref{fig:OutOfPlane}.
\begin{figure}[h]
    \centering
    \includegraphics[width=.85\linewidth, trim=0cm 0cm 0cm 0cm, clip=true]{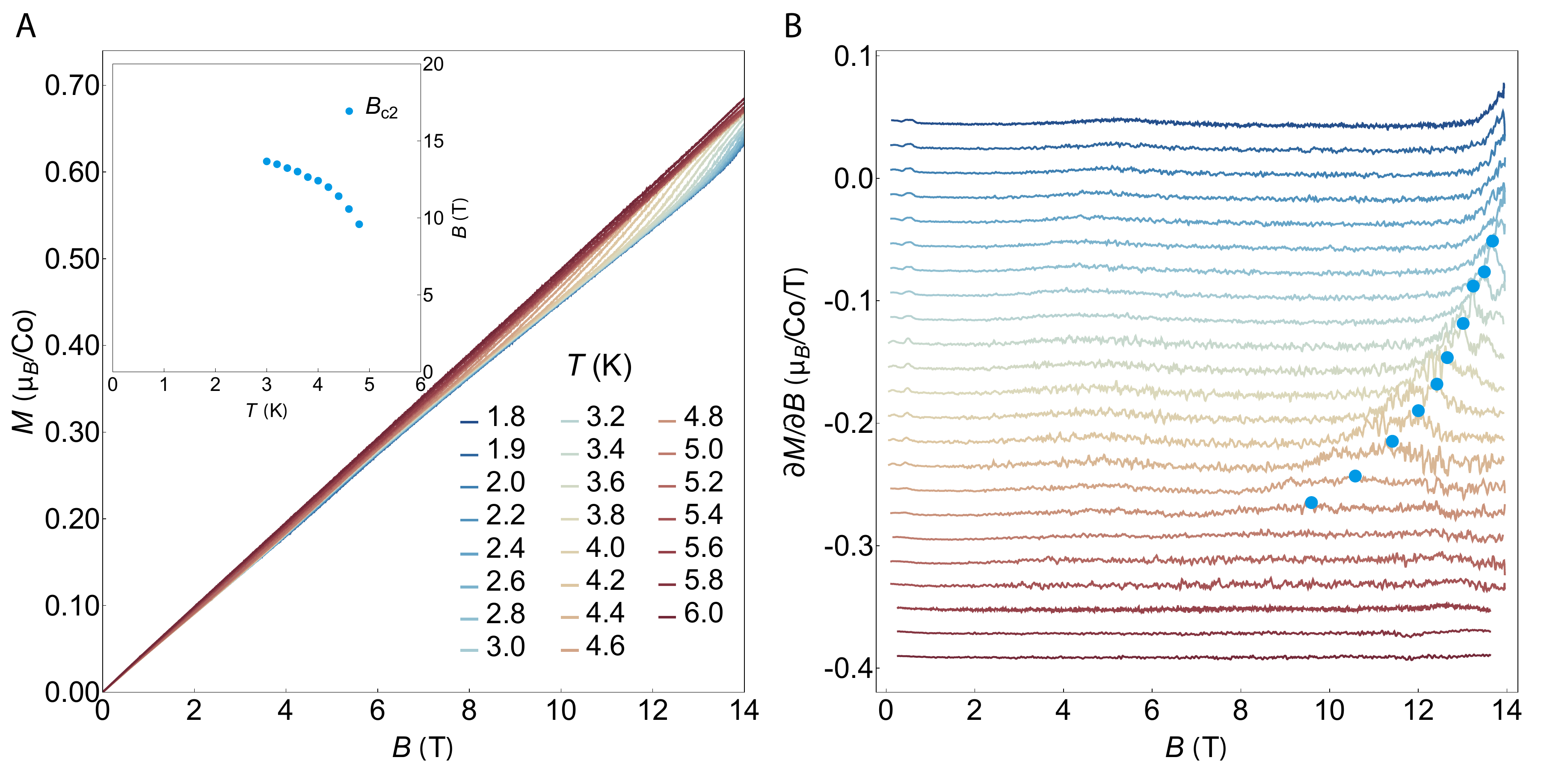}
    \rule{27em}{0.5pt}
    \caption{\scriptsize \textbf{} A) Temperature-dependent magnetization of \bcao for field applied perpendicular to the honeycomb planes ($\theta =0^\circ$). The inset shows critical field \bct as a function of temperature. B) The field-derivatives of the magnetization shown in panel A used to identify the positions of the critical field, \bct.}
    \label{fig:OutOfPlane}
\end{figure}

\subsection{Magnetotropic measurements}
\label{section:Magnetotropic}

\subsubsection{Unit conversion}

In resonant torsion magnetometry, we infer the anisotropic magnetic response of the sample by measuring a shift in the resonant frequency of a cantilever as external parameters (temperature, magnetic field, field angle) are changed. In order to have an intuitive understanding of the magnitude of the frequency shift (proportional to the magnetotropic susceptibility, $k$), we calibrate the measured response in the linear regime by using the known anisotropic magnetic susceptibility $\chi_{ii}$ and $\chi_{jj}$ at $T = 100$ K. \\

To convert the frequency shift into energy units, we first find the molar concentration of cobalt atoms in the measured sample by estimating the sample volume based on SEM images. The estimated volume of the FIB'ed sample shown in \autoref{fig:angle data} is 161 $\mu$m$^3$ and the unit cell size is 510 \AA$^3$, corresponding to $3.2 \times 10^{11}$ unit cells. Each unit cell contains 6 cobalt, leading to roughly 3.14 pmol of Co in the smaller FIB'ed sample. Estimates similarly obtained for the larger bulk crystal (upper SEM image in \autoref{fig:angle data}) give 98.4 pmol of Co.

\begin{figure}[h]
    \centering
    \includegraphics[width=0.6\textheight]{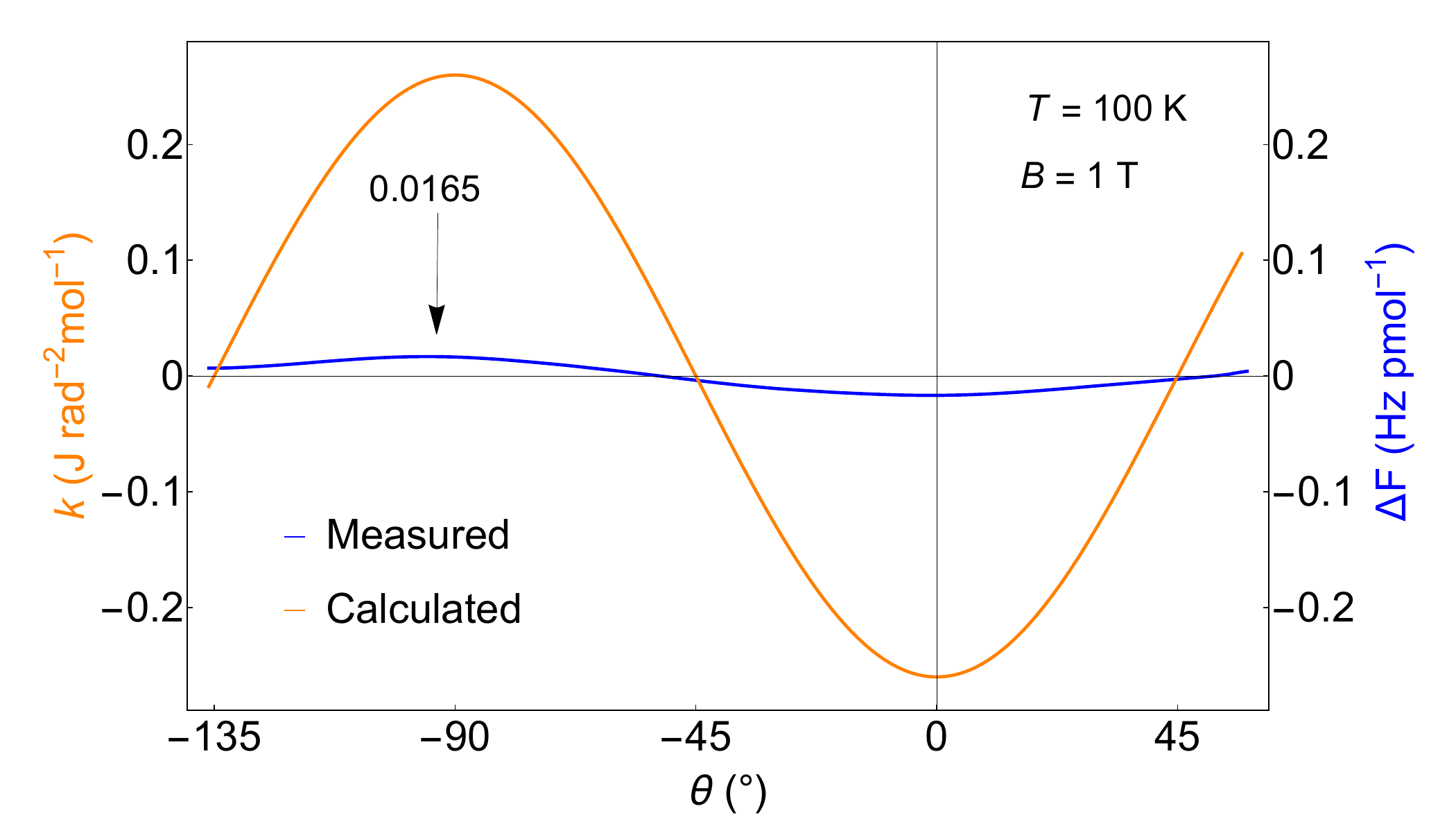}
    \caption{The magnetotropic susceptibility calculated from the known anisotropic magnetic susceptibility in the $ac$-plane \cite{Zhong_2020}. The blue curve shows the measured frequency shift at the same field and temperature as the calculated orange curve. The ratio in the amplitudes of the $\text{cos}\theta$ curves provides a scaling factor to convert the measured frequencies into energy units. We find that 1 Hz = 15.5 pJ.}
    \label{fig:Cos}
\end{figure}

\begin{figure}[h]
    \centering
    \includegraphics[width=0.6\textheight]{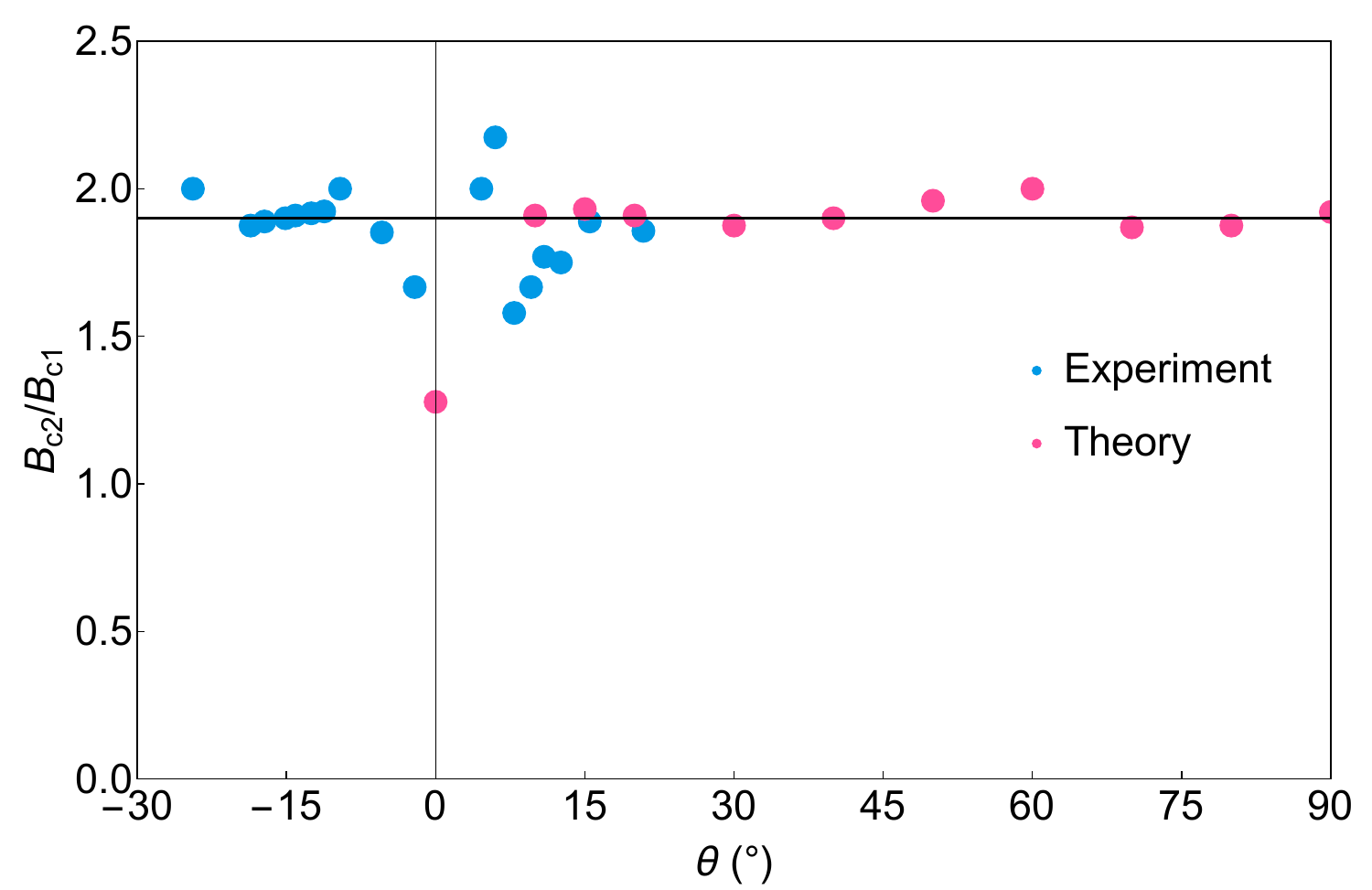}
    \caption{The experimental and calculated ratio of \bct to \bco. The solid horizontal line is a guide to the eye at \bct$/$\bco$ = 1.9$. The calculated ratio drops near $\theta = 0^{\circ}$: \bco is not visible in the experimental data at this angle. }
    \label{fig:ratios}
\end{figure}

Taking into account the relations $ k = \partial\tau/\partial \theta $,  $\tau = M \times B $, and $M = \chi H$, the magnetotropic susceptibility in the linear regime is:
\begin{equation}
    k =  (\chi_c- \chi_a) \mu_0 H^2 \text{cos}2\theta,
\end{equation}

where $\chi_a$ and $\chi_c$ represent the $a-$ and $c-$axis magnetic susceptibility, respectively.\\
Given the magnetic susceptibility values obtained from \citet{Zhong_2020}, we can calculate the magnetotropic susceptibility in units of joules per mole. \autoref{fig:Cos} shows $k$ as a function of angle at $\mu_0H$ = 1 T and $T$ = 100 K as calculated using the known magnetic susceptibility (orange curve). The blue curve shows the raw frequency shift per mole at the same field and temperature. The ratio of the orange curve to the blue curve provides the scaling factor that is then applied to all data sets.  

The FIB'ed sample (containing 3.14 pmol of cobalt atoms) shown in \autoref{fig:angle data} produces a frequency shift of $\sim$0.055 Hz at $B= 1$ T and $T = 100$ K.
The ratio of the maximum values for each curve provide a scaling factor to convert the frequency shift to joule unit. This scaling factor is 15.5 pJ per rad$^2$ per Hz.

\subsection{Monte Carlo calculations}
\label{section:MonteCarlo}

Our Monte Carlo (MC) calculations were performed using the Metropolis algorithm, employing parallel tempering and overrelaxation sweeps allowing them to converge more rapidly. The metropolis algorithm updates spins by first proposing a random change to a spin and then accepting that change with a probability $\mathrm{min}(e^{-\beta\Delta E},1)$. We perform this procedure for every spin in the simulation, thus performing a full Metropolis sweep. In addition to this type of update, we also use overrelaxation sweeps. These involve flipping a spin across the local effective field at each site
\begin{equation}
    \mathbf{S}_i\xrightarrow{}-\mathbf{S}_i+2\frac{2\mathbf{S}_i\cdot \mathbf{B}_i^\text{eff}}{|\mathbf{B}_i^\text{eff}|^2}\mathbf{B}_i^\text{eff}.
\end{equation}
These sweeps do not change the energy of the spin configuration and so allow us to more rapidly explore the space of all possible spin configurations. We perform 10 of these overrelaxation sweeps for every Metropolis sweep. Furthermore, we use parallel tempering
wherein many different temperatures are simulated simultaneously and their spin configurations are swapped with one another with probability $\mathrm{min}(e^{\Delta\beta\Delta E},1)$. We propose such a swap between adjacent temperature simulations every 20 MC sweeps. Monte Carlo calculations were performed using a $24\times24\times2$ system size. After waiting $5\times 10^6$ MC sweeps for the system to thermalize, $10^5$ measurements were taken, waiting 100 sweeps between each measurement to avoid auto-correlation. For values of $B$ and $\theta$ near the phase boundaries, the results of up to 50 separate simulations were averaged together to improve accuracy. For other points, only a single Markov chain was needed.
Magnetotropic susceptibility was computed from the measured spin configurations using Equation~\ref{eq:magnetotropic Susceptibility}.
\begin{equation}
    k=\left(\mathbf{n}\times\mathbf{B}\right)\cdot\left(\mathbf{n}\times\mathbf{M}\right)-\left(\mathbf{n}\times\mathbf{B}\right)^T\chi \left(\mathbf{n}\times\mathbf{B}\right)
    \label{eq:magnetotropic Susceptibility}
\end{equation}
where $\mathbf{n}$ is the axis about which the rotation occurs. For the MC calculations performed here, $\mathbf{B}$ was kept strictly in the $ac$-plane and therefore $\mathbf{n}$ is a unit vector perpendicular to this plane.
This is the definition of magnetotropic susceptibility comes from~\cite{Shekhter_2023}.

\begin{figure}[htbp]
    \centering
    \includegraphics[width=\linewidth, trim=1.2cm 1cm 1.2cm 2cm, clip=true]{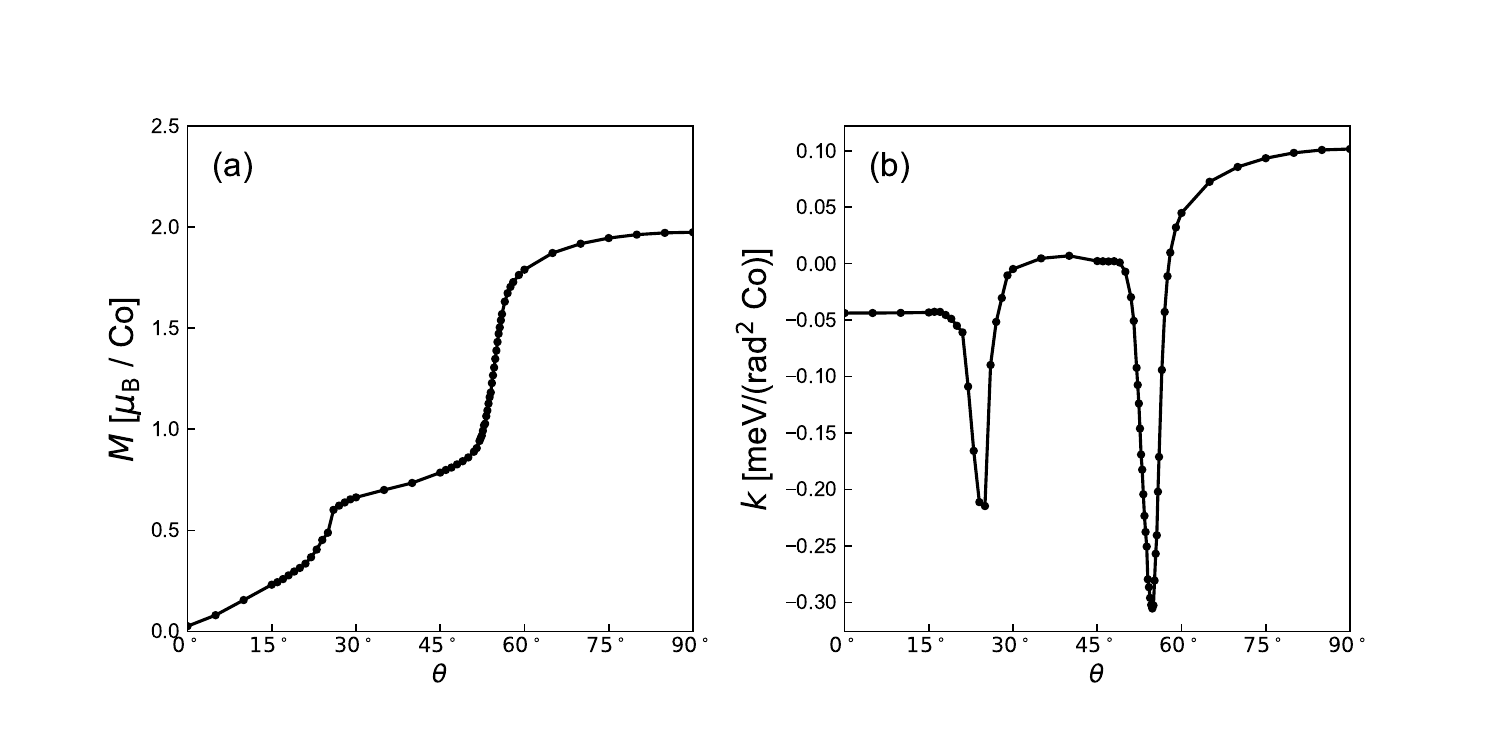}
    \caption{Results from Monte Carlo calculations shown at fixed temperature ($T=1.5$ K) and magnetic field strength ($B=0.9$ T) as a function of applied field angle $\theta$. (a) Shows magnetization per site, with the characteristic $M=1/3$ plateau corresponding to the commensurate $q=1/3$ phase. (b) Shows magnetotropic susceptibility per site as a function of $\theta$. The two peaks in $k$ coincide with the steep increase in magnetization thus we use these peaks to identify the phase boundaries.}
    \label{fig:k vs theta}
\end{figure}

\subsection{iDMRG calculations}
\label{section:iDMRG}
iDMRG was implemented using the python package TeNPy \cite{Hauschild_2018}. In iDMRG a 1D Matrix Product State (MPS) representation of the wavefunction is used to represent the 2D system by snaking around the MPS unit cell. iDMRG variationally optimizes this MPS wavefunction into the ground state. For our simulations, an XC cylindrical geometry was used. The MPS unit cell had a circumference length $L_y$ and length $L_x$ in the infinite direction. As iDMRG gives us access to long-range correlations in the infinite direction, the XC geometry permits the computation of a high-density cut of the spin structure factor between the $\Gamma$ and $M$ point in the Brillouin zone (see Figure \ref{fig:iDMRG} (a) inset).

The Hamiltonian $H_{XXZ}$ has a $U(1)$ symmetry of rotations in the $x$-$y$ plane, and this symmetry is encoded into its MPS representation. Correspondingly, $S^z_{tot}=\sum_i S^z_i$ is conserved as $\left[ S^z_{\text{tot}}, H \right]=0$. In this study, we have focused only on the $S^z_{tot}=0$ sector of the spin model.

For the iDMRG, we have followed a two-step routine. We start from an up/down product state and run a maximum of 40 sweeps with the density matrix mixer turned on to escape from any local minima in the energy landscape. Then we run a maximum of 100 sweeps or until $\Delta E=10^{-8}$ with the density matrix mixer turned off to converge into the ground state in the global minima basin. 

As shown in Figure \ref{fig:iDMRG}, we have performed our iDMRG simulations for different MPS unit cell sizes $(L_x, L_y)$, and different bond dimensions $\chi$. Even though the phase boundaries weakly depend on these choices due to finite size effects from wrapping around a cylinder and the truncation of less relevant entanglement information, the overall structure of the phase diagram is unaffected.




For a consistency check with the finite slab DMRG simulations in \cite{Jiang_2023}, we used the same parameter set for the XXZ-$J_1$-$J_3$ model without the bond-dependent anisotropies in an iDMRG simulation. The parameter set from \cite{Jiang_2023} corresponds to $J_{xy}^{(1)}=-1.00$, $J_{xy}^{(1)}/J_{xy}^{(3)}=-3.00$, $J_{z}^{(1)}/J_{xy}^{(1)}=0.13$ and $J_{z}^{(3)}/J_{xy}^{(3)}=-0.35$. Even without the bond-dependent anisotropies, we find a ferromagnetic ground state, this is in agreement with the results of \cite{Jiang_2023}. However, the ratios $J_{z}^{(1)}/J_{xy}^{(1)}$ and $J_{z}^{(3)}/J_{xy}^{(3)}$ used in the parameter set from \cite{Jiang_2023} do not match with Eq 13 of Ref. \cite{Halloran_2022} or our more refined parameter set in \autoref{eq:classical_parameters}.

\subsection{Model Parameters in Monte Carlo and iDMRG simulations} 
\label{section:Model Parameters}

The parameters $D_\lambda$ and $E_\lambda$ in Eq.~\ref{eq:Hamiltonian} are explicitly
\begin{equation}
    \begin{split}
    D_\lambda&=D\cos(2\alpha_\lambda)-E\sin(2\alpha_\lambda)\\
    E_\lambda&=D\sin(2\alpha_\lambda)+E\cos(2\alpha_\lambda)
    \end{split}
\end{equation}
where $\alpha_\lambda=+2\pi/3, -2\pi/3,0$ for $\lambda=x,y,z$ respectively.
The values of parameters used for all MC calculations are 
\begin{equation}
    \begin{array}{cc}       J_{xy}^{(1)}=-7.65\mathrm{meV} & J_{xy}^{(3)}=+2.64\mathrm{meV}\\
    J_{z}^{(1)}=-1.20\mathrm{meV} & J_{z}^{(3)}=-0.81\mathrm{meV}\\    D=+0.10\mathrm{meV} & E=-0.10\mathrm{meV}\\
    g_{ab}=5.0 & g_c=2.7\\
    \end{array}\label{eq:classical_parameters}
\end{equation}
these are a refined version of the parameter set in \cite{Halloran_2022}, modified slightly to improve agreement with experimentally observed critical fields. For the iDMRG simulations shown in \autoref{fig:iDMRG} (a), we have used $J_{z}^{(1)}=-1.20$ meV and $J_{z}^{(3)}=-0.80$ meV. Further, we have used the constraint $J_{xy}^{(1)}+J_{xy}^{(3)}=-5.00$ meV to compare with the classical phase diagram presented in Figure 6 e in Ref. \cite{Halloran_2022}.

\section*{Acknowledgments}
We are very grateful for the helpful discussions with Sasha Chernyshev, Tom Halloran, Collin Broholm, and Peter Armitage. S.S., Z.K., and K.A.M. acknowledge support from the collaborative research project SFB Q-M\&S funded by the Austrian Science Fund (FWF) F 86 and ISTA. S.S., Z.K., and K.A.M. thank the ISTA Nanofabrication Facility for technical support. B. J. R., A. M. F., and T. M. M. acknowledge funding from the Institute for Quantum Matter, an Energy Frontier Research Center funded by the Office of Basic Energy Sciences of the United States Department of Energy under Award No. DE-SC0019331 (sample growth and preparation, sample characterization). T.M.M. acknowledges support of the David and Lucile Packard Foundation. Y.B.K., W.B.H., A.M., F.D., and E.Z.Z. are supported by the NSERC (Natural Science and Engineering Council of Canada) Discovery Grant No.~RGPIN-2023-03296 and the Center for Quantum Materials at the Univeristy of Toronto. A.S. acknowledges support at Los Alamos National Laboratory by the NSF through DMR-1644779 and DMR-2128556 and the U.S. Department of Energy. A.S. acknowledges support from the DOE/BES ``Science of 100 T" grant.

%


\end{document}